\documentclass[conference]{IEEEtran}
\IEEEoverridecommandlockouts

\usepackage{cite}
\usepackage{amsmath,amssymb,amsfonts}
\usepackage{algorithmic}
\usepackage{graphicx}
\usepackage{subcaption}
\usepackage{textcomp}
\usepackage{xcolor}
\usepackage{url}
\usepackage{hyperref}
\def\BibTeX{{\rm B\kern-.05em{\sc i\kern-.025em b}\kern-.08em
    T\kern-.1667em\lower.7ex\hbox{E}\kern-.125emX}}
\begin{document}

\title{MedBike: A Cardiac Patient Monitoring System Enhanced through Gamification}
\vspace{-5mm}

\author{\IEEEauthorblockN{Tahmim Hossain}
\IEEEauthorblockA{\textit{Department of Computing Science} \\
\textit{University of Alberta}\\
Edmonton, Canada \\
tahmim@ualberta.ca}
\and
\IEEEauthorblockN{Faisal Sayed}
\IEEEauthorblockA{\textit{Department of Computing Science} \\
\textit{University of Alberta}\\
Edmonton, Canada \\
faisalz1@ualberta.ca}
\and
\IEEEauthorblockN{Yugesh Rai}
\IEEEauthorblockA{\textit{Department of Computing Science} \\
\textit{University of Alberta}\\
Edmonton, Canada \\
yugesh@ualberta.ca}

\and
\hspace{35mm}\IEEEauthorblockN{Kalpak Bansod}
\IEEEauthorblockA{\hspace{35mm}\textit{Department of Computing Science} \\
\hspace{35mm}\textit{University of Alberta}\\
\hspace{35mm} Edmonton, Canada \\
\hspace{35mm} kalpakni@ualberta.ca}

\and
\IEEEauthorblockN{Md Nahid Sadik}
\IEEEauthorblockA{\textit{Department of Computing Science} \\
\textit{University of Alberta}\\
Edmonton, Canada \\
msadik@ualberta.ca}
}

\maketitle

\begin{abstract}
The "MedBike" is an innovative project in the field of pediatric cardiac rehabilitation. It is a 2D interactive game created specifically for children under the age of 18 who have cardiac conditions. This game is part of the MedBike system, a novel rehabilitation tool combining physical exercise with the spirit of gaming. The MedBike game provides children with a safe, controlled, and engaging environment in which to exercise and recover. It has three distinct levels of increasing intensity, each with its own set of environments and challenges that are tailored to different stages of rehabilitation. This report dives into the details of the MedBike game, highlighting its unique features and gameplay.
\end{abstract}


\section{Introduction and Background}

The MedBike project, a pioneering initiative in pediatric cardiac rehabilitation, has played an important role in increasing patient interest in rehabilitation programs. It was created by Advanced Human-Computer Interfaces (AHCI) Labs with a dual-interface design that caters to both patients and clinicians, allowing children to participate in rehabilitative exercise from the comfort of their own homes. This system, which includes vital health monitoring tools such as ECG, arterial oxygen saturation, and blood pressure sensors, provides a safe and controlled exercise environment. The clinician's ability to monitor up to six patients simultaneously, with live access to biometric data and a bidirectional video and audio feed, represents a significant advance in remote patient care. By addressing common barriers such as distance, motivation, and social support, the MedBike system has significantly contributed to lowering the incidence of mortality and disease.

MedBike has been a valuable tool in cardiac rehabilitation since its inception, particularly for paediatric patients. The system's design, which allows patients to cycle through different environments, has been critical in keeping patients engaged. Despite its initial success, the MedBike system encountered difficulties that required us to create a new game for the MedBike systems. As per the feedback from the Client , University of Alberta Health Research Institute of MedBike systems who have been actively using it with the children with caridiac problems, we identified the following need for update/features to be included in the new game:

\begin{enumerate}
    \item \textbf{Need for New Exciting Levels:} The existing game within the MedBike system had not been updated since it's first release. This lack of updates led to a stagnation in the user experience, with the game no longer providing the novelty and excitement necessary to keep young patients engaged.
\item \textbf{Technical Glitches in Existing Levels: } Technical issues in the game, such as the absence of collision detection in few existing levels that allowed users to pass through solid objects like lakes,trees etc detracted from the realism and immersion of the experience. This flaw in the game's physics engine not only reduced the quality of the gameplay but also potentially impacted the therapeutic effectiveness of the exercise.
   \item \textbf{ Enhancement of Audio Experience:} The music and sound effects within the game were identified as monotonous and uninspiring. For a demographic as sensitive to engagement as children, the auditory experience plays a crucial role in maintaining interest and motivation. Revamping the game's audio component was essential to reinvigorate the user experience.
\item \textbf{Overall System Refresh:} Since the existing system had not received any updates since it's first development, the entire system required a refresh to stay relevant and effective. This refresh needed to encompass both the visual appeal and the functional aspects of the game, ensuring that it remains a compelling part of the rehabilitation process.

\end{enumerate}

In response to these identified needs, our team, as part of the MM 806 course project, decided to create a new version of the MedBike game. This initiative was driven by the goal of enhancing the MedBike experience, making it more immersive, engaging, and technically robust.

\section{Client Requirement}

\begin{itemize}
    \item Develop a fully functional game with high intensity and cool down levels
    \item Ensure the game is engaging and suitable for children with cardiac conditions
    \item Make the game for PC so it can be integrated
    \item Conduct usability testing to ensure safety and effectiveness.
\end{itemize}

Our team closely adhered to the specific requirements and feedback provided by our client, the University of Alberta Health Research Institute, when developing the new version of the MedBike game. We had originally planned an ambitious improvement of the game, including Virtual Reality (VR) version for a more immersive experience. This plan, however, was significantly altered as a result of advice from our client and course professor, who emphasised the importance of safety and feasibility. The primary directive was to shift our focus to developing a 2D version of the game, given the end-users' cardiac conditions and the safety concerns associated with more immersive VR environments. This decision was also consistent with the existing capabilities of the MedBike system, which are optimised for 2D displays.
We focused on the core development of the game using the Unity engine for 2D gameplay when refining the scope of our project, a decision influenced by the tight project deadline. This focus enabled us to effectively improve the game's design, user experience, and technical features within the time constraints. In terms of content, the client requested the creation of three levels and two black holes for this version, with the goal of striking a balance between engagement and simplicity. To maintain a child-friendly and psychologically appropriate environment for our young audience, an initial concept to include 'scary' elements in the game was reevaluated and ultimately discarded. 
Furthermore, integration of the game with the existing bicycle system, which was a key component of the MedBike experience, was not pursued. The professor's advice and the project's time constraints influenced our decision to focus solely on standalone game development. The client's willingness to consider future game expansions based on user feedback also influenced our development process significantly. This openness implies a dedication to the game's continuous improvement and adaptation, ensuring its long-term relevance in paediatric cardiac rehabilitation. Our collaborative approach, which was closely aligned with the client's needs and constraints, resulted in the development of a safe, engaging, and technically sound 2D game.

\section{Methodology}

\begin{itemize}
    \item Project selection
    \item Meeting up with client to understand about the project and the requirements.
    \item Acquiring existing data and problems of the project.
    \item Visualizing the project.
    \item Implementing the visualized idea to develop our project.
    \item Presenting a demo of our project to the client.\\
\end{itemize}

This report provides a succinct overview of the procedural workflow inherent to our gaming model. A detailed exploration of the sequential steps involved in launching and navigating through the application is presented, offering a comprehensive understanding of the user experience.\\

\begin{figure}[htbp]
  \centering
  \includegraphics[width=0.386\textwidth]{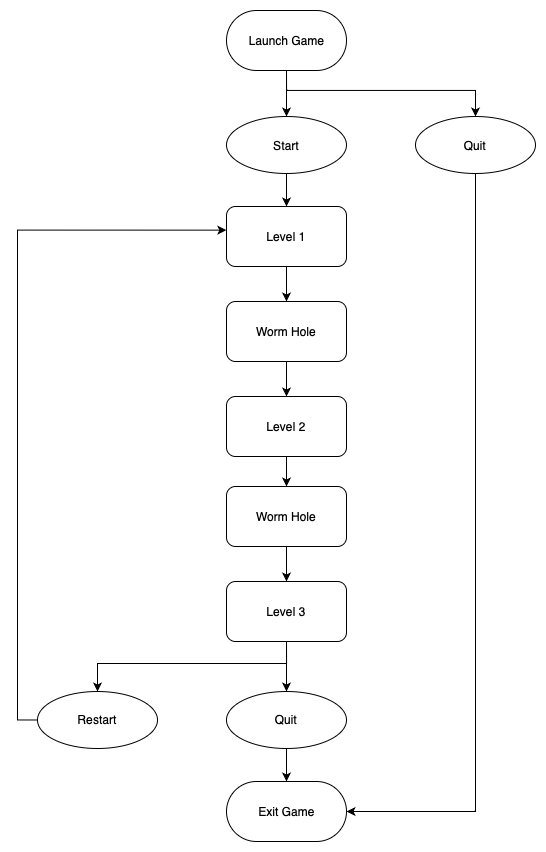}
  \caption{Workflow}
\end{figure}

\textbf{\textit{Workflow:}}

\textbf{Game Launch:}
The initiation of the model transpires through the act of clicking on the compiled game file. Upon execution, the game promptly loads, unveiling the initial landing screen.

\textbf{Landing Screen:}
The landing screen serves as the primary interface for users, presenting them with two discernible options: 'Start' and 'Quit.' In the event of selecting 'Quit,' the game gracefully terminates, bringing the application to a conclusion. Conversely, opting for 'Start' propels users into the immersive gaming experience.

\textbf{Level One:}
The gaming journey commences with Level One, setting the stage for the subsequent challenges.

\textbf{Wormhole:}
Following the completion of Level One, users are seamlessly transported into a dynamic and challenging Wormhole segment, where precision and speed are paramount.

\textbf{Level Two:}
The gameplay advances to Level Two, introducing fresh challenges and scenic environments for users to navigate.

\textbf{Subsequent Wormhole:}
A second encounter with the Wormhole follows Level Two, maintaining the intensity and demanding user engagement.

\textbf{Level Three:}
The pinnacle of the gaming experience unfolds in Level Three, offering a crescendo of challenges and culminating in a fulfilling gaming journey.

\textbf{Restart Screen:}
Upon the triumphant completion of Level Three, users are directed to a Restart Menu, providing them with two choices: 'Quit' or 'Restart.'
Restart Option:
Opting for 'Restart' seamlessly reloads Level One, allowing users to relish the gaming experience afresh.
Quit Option:
Choosing 'Quit' concludes the gaming session and gracefully terminates the application.

\section{Softwares \& Assetd}

\textbf{Unity 3D:}
Unity 3D was the main software we used for developing our game all the effects, physics, scripting, and implementation of (C Sharp) code to introduce logic to our game and add functionalities. It is basically the platform on which the project is developed. Unity 3D allows developers to develop games for various platforms, including PC, Mac, iOS, Android, and more. Unity 3D is also used for architectural visualizations, training simulations, and other types of interactive media.\\

\textbf{Adobe Illustrator:}
We used Adobe Illustrator to develop our homepage and final page, we developed buttons (start, quit, restart) and made some design improvements using this software. Adobe Illustrator is the industry-leading graphic design tool that lets you design anything you can imagine – from logos and icons to graphics and illustrations – and customize it with professional-level precision, as well as time-saving features like Repeat for Patterns or Global Edits. We created a variety of graphics produced logos and detailed illustrations, required in our project.\\
 
\textbf{Blender:}
Some of the 3d objects like trees, rocks, mountaineer plain, and many more, were created and imported it into our game using Blender. Blender is an open-source 3D computer graphics software toolset used for creating animated films, visual effects, art, 3D-printed models, motion graphics, interactive 3D applications, virtual reality, and, formerly, video games. Blender's features include 3D modeling, UV mapping, texturing, digital drawing, raster graphics editing, rigging and skinning, fluid and smoke simulation, particle simulation, soft body simulation, sculpting, animation, match moving, rendering, motion graphics, video editing, and compositing.\\

As the assets, we mainly used 4. `Simple Bicycle Physics`\cite{AiKodex_2023} is used for the Bicycle and the physics of the bicycle. `Cartoon Racetrack Oval`\cite{Design_2021} is used for the first level. `Mountain Racetrack`\cite{DIGITALS_2019} is used as the second level. Finally `Lunar Landscape 3D`\cite{Squirrel_2019} is used as the final level. The wormhole level is designed from scratch.\\

\section{System Design}
Our MedBike application is meticulously designed to provide users with an engaging and health-oriented experience. The following report outlines the fundamental structure and layout of the game, detailing its menu pages, basic levels, and high-intensity challenges.

\textbf{\textit{Structure and Layout:}}

\textbf{Basic Levels:}
Our gameplay is categorized into Basic Levels (1, 2, and 3) and High-Intensity Levels (Wormhole). Basic Levels offer a serene and leisurely cycling experience, each spanning a duration of three minutes. The picturesque pathways are characterized by twists, turns, inclinations, and declines, providing users with a captivating environment. Level one unfolds on a racetrack, level two amidst a scenic forest mountain setting, and level three in a captivating space-themed environment, adorned by the moon.

\begin{figure}[htbp]
  \centering
  \includegraphics[width=0.386\textwidth]{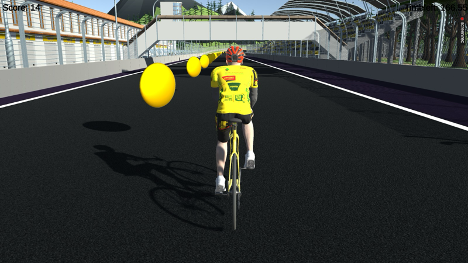}
  \caption{Level 1}
\end{figure}
\begin{figure}[htbp]
  \centering
  \includegraphics[width=0.386\textwidth]{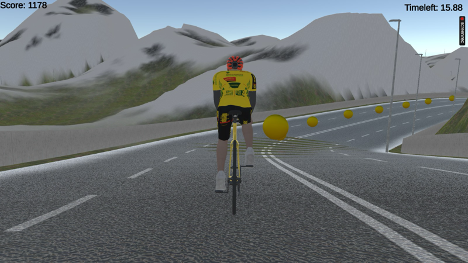}
  \caption{Level 2}
\end{figure}
\begin{figure}[htbp]
  \centering
  \includegraphics[width=0.386\textwidth]{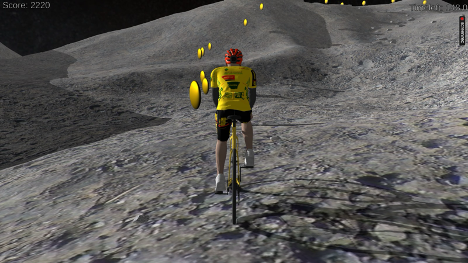}
  \caption{Level 3}
\end{figure}

\textbf{High-Intensity Levels (Wormhole):}
The Wormhole, a pinnacle of high-intensity challenges, maintains a consistent layout with Basic Levels. Users are actively encouraged to exert maximum effort, reaching their peak cycling speed. A straight pathway facilitates the achievement of top speed, and users are duly rewarded with an exhilarating 3X points multiplier. The captivating space-themed backdrop, complete with stars gracefully passing by, intensifies the challenge and motivates users to push their physical limits.

\begin{figure}[htbp]
  \centering
  \includegraphics[width=0.386\textwidth]{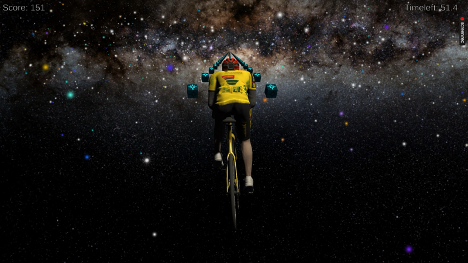}
  \caption{Wormhole}
\end{figure}

\textbf{Layout of Game Levels:}
The user commands the central focus, positioned at the center of the screen while cycling alongside the pathway the pickup prefabs are placed for the user to collect in order to gain points. Essential information such as the remaining time is elegantly displayed in the top right corner. Simultaneously, the user's score is showcased in the top left corner. As the timer approaches three seconds, a strategically placed pop-up message emerges at the center of the screen, heralding the imminent 'Level Up' moment. To augment the user experience, distinct background music complements each level, accompanied by subtle sounds signifying point collection.

\textbf{\textit{Menu Pages:}}

\textbf{Landing Menu:}
serves as the gateway to the gaming experience, featuring the title prominently. Users are presented with two options: 'Play' and 'Quit.' The 'Play' option initiates the gameplay, while 'Quit' concludes the application.
\begin{figure}[htbp]
  \centering
  \includegraphics[width=0.386\textwidth]{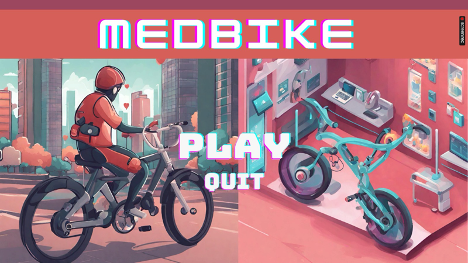}
  \caption{Landing Menu}
\end{figure}

\textbf{Restart Menu:}
Upon the triumphant completion of the third level, the seamless transition to the restart menu occurs. Users are presented with the option to either relish the immersive experience once more by selecting 'Restart' or gracefully exit the application by choosing 'Quit.' Apart from the ‘replay and Quit’ option the Restart screen also displays the total score of the player.

\begin{figure}[htbp]
  \centering
  \includegraphics[width=0.386\textwidth]{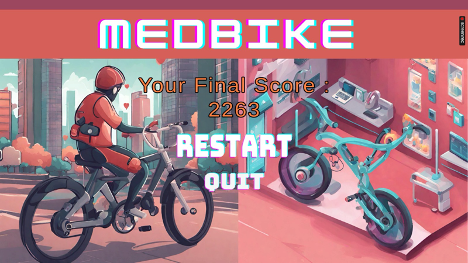}
  \caption{Restart Menu}
\end{figure}

This meticulously crafted structural framework and gameplay sequence amalgamate entertainment with health consciousness, offering users an engaging yet physically rewarding cycling experience.\\

\textbf{GamePlay:}

\begin{figure}[htbp]
  \centering
  \includegraphics[width=0.386\textwidth]{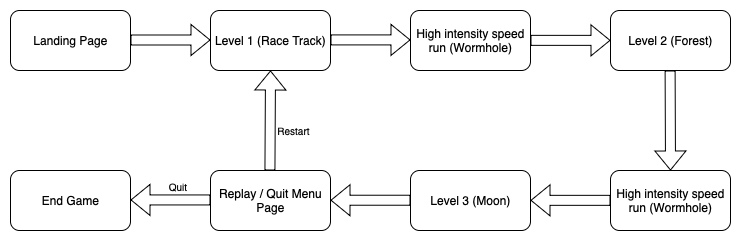}
  \caption{Chronology}
\end{figure}

\section{Contributions}

The development of the new MedBike game was a collaborative effort involving various tasks, each critical to the project's success. The team members contributed in different capacities, ensuring a well-rounded and effective development process.

\textbf{Project Planning and Research:} All team members were involved in defining specific game requirements and researching pediatric cardiac rehabilitation guidelines. This included understanding the needs of the target demographic and aligning the game's features with therapeutic goals.

\textbf{Unity 3D Development and Landing Page:} Nahid and Tahmim set up the Unity project and developed the regular-level environment and gameplay. They focused on creating a user-friendly interface and ensuring smooth gameplay mechanics.

\textbf{Creating 3 Levels for the Game:} Yugesh and Kaplak designed the Race Track, Mountain track, and Moon (space) environment. They worked on creating diverse and engaging levels that would keep the players interested and motivated.

\textbf{Creating Black Hole Level:} Kalpak and Faisal created the black hole level and conducted testing and optimization. 

\textbf{Creating Prefabs and Materials:} Faisal and Nahid were responsible for creating prefabs like coins, stars, materials, etc., for each level. They ensured that these elements were visually appealing and functionally sound.

\textbf{Integrating Levels and Logic:} Yugesh, Tahmim, and Kalpak worked on integrating all levels, including the implementation of score count and timer logic. They focused on ensuring a seamless transition between levels and maintaining consistent gameplay.

\textbf{Testing and Usability Survey:} Faisal, Yugesh, and Tahmim conducted comprehensive testing and debugging, deployment, and gathered usability feedback from users. They aimed to identify and fix any issues to enhance the overall user experience.

Every member was involved in most of the tasks, ensuring a consistent level of oversight and contribution across the project. Each member played a vital role in their respective areas, contributing to the project's overall success.

\section{Usability Study Results}
The usability study for the MedBike application yielded insights from 21 participants, reflecting diverse preferences in bike workout routines. The majority of users (66.7\%) favored high-intensity interval training, while others engaged in leisure cycling (23.8\%), Cyclothon (4.8\%), or cycle races (4.8\%). Installation proved seamless for 81\% of participants, with only 19\% encountering issues. Notably, 95.2\% found the time duration of levels ideal, and user performance in the high-intensity Wormhole level received positive ratings. Background sound, rated on a scale of 1 to 5, was generally well-received, with 61.9\% giving the highest rating. The overall application experience earned high marks, with 71.4\% providing a maximum rating. Participants suggested improvements, including the incorporation of psychedelic music, additional levels and quests, increased level complexity, more bike skin options, and customizable background music preferences. Some users expressed satisfaction with the current state of the application, emphasizing its clear integration and novel features. These findings offer valuable insights for refining and enhancing the MedBike application to better cater to user preferences and expectations.

\begin{figure}[htbp]
  \centering
  \includegraphics[width=0.386\textwidth]{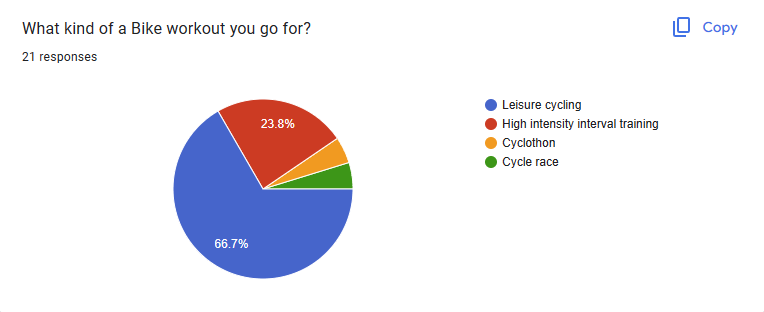}
  \caption{Bike Workout preference}
\end{figure}
\begin{figure}[htbp]
  \centering
  \includegraphics[width=0.386\textwidth]{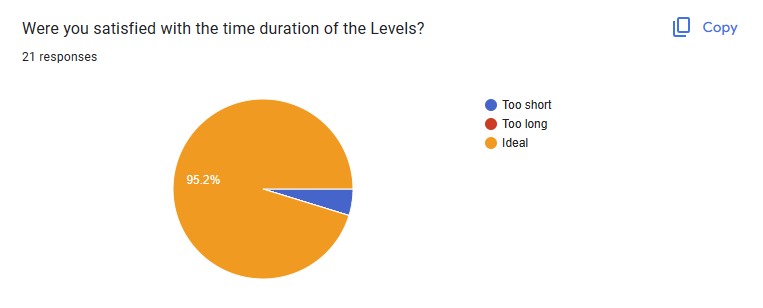}
  \caption{Level Duration Preferennce}
\end{figure}
\begin{figure}[htbp]
  \centering
  \includegraphics[width=0.386\textwidth]{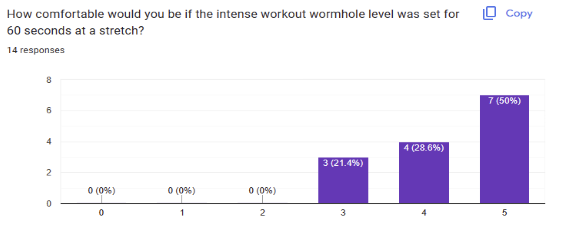}
  \caption{Wormhole intensity}
\end{figure}

\section{Conclusion and Future Scope}
In conclusion, the "MedBike" project stands as an advancement in the realm of pediatric cardiac rehabilitation. Tailored specifically for children under the age of 18 with cardiac conditions, the 2D interactive game within the MedBike system seamlessly combines physical exercise with the immersive spirit of gaming. The incorporation of three distinct levels of increasing intensity demonstrates the thoughtful design, ensuring that the game aligns with the different stages of rehabilitation. Each level presents a set of environments and challenges that not only cater to the diverse needs of patients but also contribute to a holistic and enjoyable rehabilitation experience.\\

Based on the user feedback, the future scope for MedBike involves implementing several key enhancements. Firstly, a robust customization feature will be integrated, allowing users to personalize their experience by selecting background music that resonates with their preferences. The addition of a variety of bike skins will cater to users seeking visual diversity, promoting a more engaging and enjoyable simulation. To elevate user engagement, a quest or milestone system will be introduced, providing users with objectives and a sense of achievement as they progress through the application. Lastly, the inclusion of more levels featuring intense workouts will cater to users seeking a higher fitness challenge, ensuring that MedBike remains a versatile and adaptive platform for individuals with varying fitness levels.\\

\bibliographystyle{IEEEtran}
\bibliography{conference_101719}
\vspace{12pt}

\end{document}